\documentclass[runningheads]{llncs}
\usepackage{graphicx}

\usepackage[normalem]{ulem} 

\usepackage{hyperref}
\hypersetup{
	colorlinks=true,       
	linkcolor=blue,          
	citecolor=blue,        
	filecolor=blue,      
	urlcolor=blue           
}

\begin{document}
\title{Post-Brexit power of European Union \\ from the world trade network analysis}
%
%
\author{Justin Loye\inst{1,2} \and
Katia Jaffr\`es-Runser\inst{3}
\and
Dima L. Shepelyansky\inst{2}
}
\authorrunning{J.Loye et al.}
%
\institute{Institut de Recherche en Informatique de Toulouse, \\
Universit\'e de Toulouse, UPS, 31062 Toulouse, France\\
\email{justin.loye@irit.fr}\\
\and
Laboratoire de Physique Th\'eorique, 
Universit\'e de Toulouse, CNRS, UPS, 31062 Toulouse, France\\
\email{dima@irsamc.ups-tlse.fr}\\
\and
Institut de Recherche en Informatique de Toulouse, \\
Universit\'e de Toulouse, Toulouse INP, 31071 Toulouse, France\\
\email{katia.jaffresrunser@irit.fr}\\
}
\maketitle              
\begin{abstract}
We develop the Google matrix analysis of the multiproduct world trade network
obtained from the  UN COMTRADE database in recent years.
The comparison is done between this new approach and the usual
Import-Export description of this world trade network. 
The Google matrix analysis takes into account the 
multiplicity of trade transactions thus highlighting in a better way
the world influence of specific countries and products.
It shows that after Brexit, the European Union of 27 countries
has the leading position in the world trade network ranking, being
ahead of USA and China.  Our approach determines also a sensitivity of
trade country balance to specific products showing the dominant role of 
machinery and mineral fuels in multiproduct exchanges. 
It also underlines the growing influence of Asian countries.

\end{abstract}

\setlength{\tabcolsep}{3pt}

\section{Introduction}
%
%
%
%
%
%
The European Union (EU) is now composed from 27 countries 
and is considered as a major world leading power \cite{eusite}.
January 2021 has seen Brexit officially taking place, triggering the  
withdrawal of the United Kingdom (UK) from EU \cite{wikibrexit}.
This event has important political, economical and social effects.
Here we project and study its consequences from the view point
of international trade between world countries.
Our analysis is based on the UN COMTRADE database \cite{comtrade} for
the multiproduct trade between world countries in 
recent years. From this database we construct the world trade
network (WTN) and evaluate the influence and trade power 
of specific countries using the  Google matrix analysis
of the WTN. We consider 27 EU countries as  a single
trade player having the trade exchange between EU and other countries.
Our approach uses the Google matrix tools and algorithms developed for the WTN 
\cite{wtn1,wtn2,wtn3,keu9} and other complex directed networks
\cite{rmp2015,politwiki}. The efficiency of the Google matrix and PageRank algorithms
is well known from the World Wide Web network analysis \cite{brin,meyer}.

Our study shows that the Google matrix approach (GMA) allows to characterize 
in a more profound manner the trade power of countries 
compared to the usual method relying on import and export analysis (IEA) between countries.
GMA's deeper analysis power originates in the fact that it accounts for the multiplicity
of transactions between countries while IEA only takes into account the 
effect of one step (direct link or relation) transactions. In this paper, we show that the 
world trade network analysis with GMA identifies EU as the first trade player in the world, 
well ahead of USA and China.

This paper is structured in the following way. Section~\ref{sec:dataset} introduces first the UN COMTRADE dataset, 
and then gives a primer on the tools related to Google matrix analysis such as the trade balance metric and the REGOMAX algorithm.
In Section~\ref{sec:results}, the central results of this papers are presented, which are discussed in \ref{sec:discussion}. 

\section{Data sets, algorithms and methods}\label{sec:dataset}

We use the UN COMTRADE data \cite{comtrade} for years 2012, 2014, 2016 and 2018 to 
construct the trade flows of the multiproduct WTN following the procedure
detailed in \cite{wtn2,wtn3}. This paper gives the results for year 2018 only, others are 
to be found at \cite{ourwebpage}. Each year is presented by a money matrix, 
$M^{p}_{cc^{\prime}}$, giving the export flow of product $p$
from country $c^{\prime}$ to country $c$ (transactions are expressed in USD of current year).
The data set is given by $N_{c} = 168$ countries and territories 
(27 EU countries are considered as one country) and $N_{p} = 10$ principal type  of products
(see the lists in \cite{wtn1,wtn3}). These 10 products are: Food and live animals (0);
Beverages and tobacco (1); Crude materials, inedible, except fuels (2);  Mineral fuels etc (3);
Animal and vegetable oils and fats (4); Chemicals and related products, n.e.s. (5);
Basic manufactures (6); Machinery, transport equipment (7); 
Miscellaneous manufactured articles (8);
Goods not classified elsewhere (9) (product index $p$ is given in brackets). They belong to 
the Standard International Trade Classification (SITC Rev. 1)
Thus the total Google matrix $G$ size is given by all system nodes $N=N_c N_p= 1680$
including countries and products.

The Google matrix $G_{ij}$ of direct trade flows
is constructed in a standard way described in detail at \cite{wtn2,wtn3}: monetary trade flows 
from a node $j$ to node $i$  are normalized to unity 
for each column $j$ thus given the matrix $S$ of 
Markov transitions for trade, the columns of dangling nodes with
zero transactions are replaced by a column with all elements being $1/N$. 
The weight of each product is taken into account  via a
certain personalized vector taking into account 
the weight of each product in the global trade volume.
We use the damping factor $\alpha=0.5$. 
The Google matrix is $G_{ij}=\alpha S_{ij} + (1-\alpha) v_i$
where $v_i$ are components of positive column vectors called personalization vectors
which take into account the weight of each product in the global trade 
($\sum_i v_i=1$).
We also construct the matrix $G^*$ for the inverted trade flows.

The stationary probability distribution 
described by $G$ is given by the PageRank vector
$P$ with maximal eigenvalue $\lambda=1$:
$GP=\lambda P =P$ \cite{rmp2015,brin,meyer}. 
In a similar way, for   the inverted flow, described by $G^*$, 
we have the CheiRank vector $P^*$, 
being the eigenvector of $G^* P^* = P^*$. 
 PageRank $K$ and CheiRank $K^*$ indexes are obtained from
monotonic ordering of probabilities of PageRank vector $P$ and 
of CheiRank vector $P^*$ as
$P(K)\ge P(K+1)$ and $P^*(K^*)\ge P^*(K^*+1)$ with $K,K^*=1,\ldots,N$.
The sums over all products $p$ gives the PageRank and CheiRank 
probabilities of a given country as $P_c =\sum_p P_{cp}$ and 
${P^*}_c =\sum_p {P^*}_{cp}$ (and in a similar way
product probabilities $P_p, {P^*}_p$) \cite{wtn2,wtn3}.
Thus with these probabilities we obtain the related indexes $K_c, {K^*}_c$.
We also define 
from import and export trade volume the
 probabilities
$\hat{P}_p$, $\hat{P}^*_p$, $\hat{P}_c$, $\hat{P}^*_c$,
$\hat{P}_{pc}$, $\hat{P}^*_{pc}$ and 
corresponding indexes
$\hat{K}_p$, $\hat{K}^*_p$, $\hat{K}_c$, $\hat{K}^*_c$, $\hat{K}$, $\hat{K}^*$
(these import and export probabilities
are normalized to  unity
by the total import and export volumes, see details in
\cite{wtn2,wtn3}). It is useful to note that qualitatively
PageRank probability is proportional to the volume of ingoing
trade flow and CheiRank respectively to outgoing flow.
Thus, we can approximately consider that the
high import gives a high  PageRank $P$ probability
and a high export a high CheiRank $P^*$ probability.

\begin{figure}[!ht]
	\centering
	\includegraphics[width=0.99\textwidth]{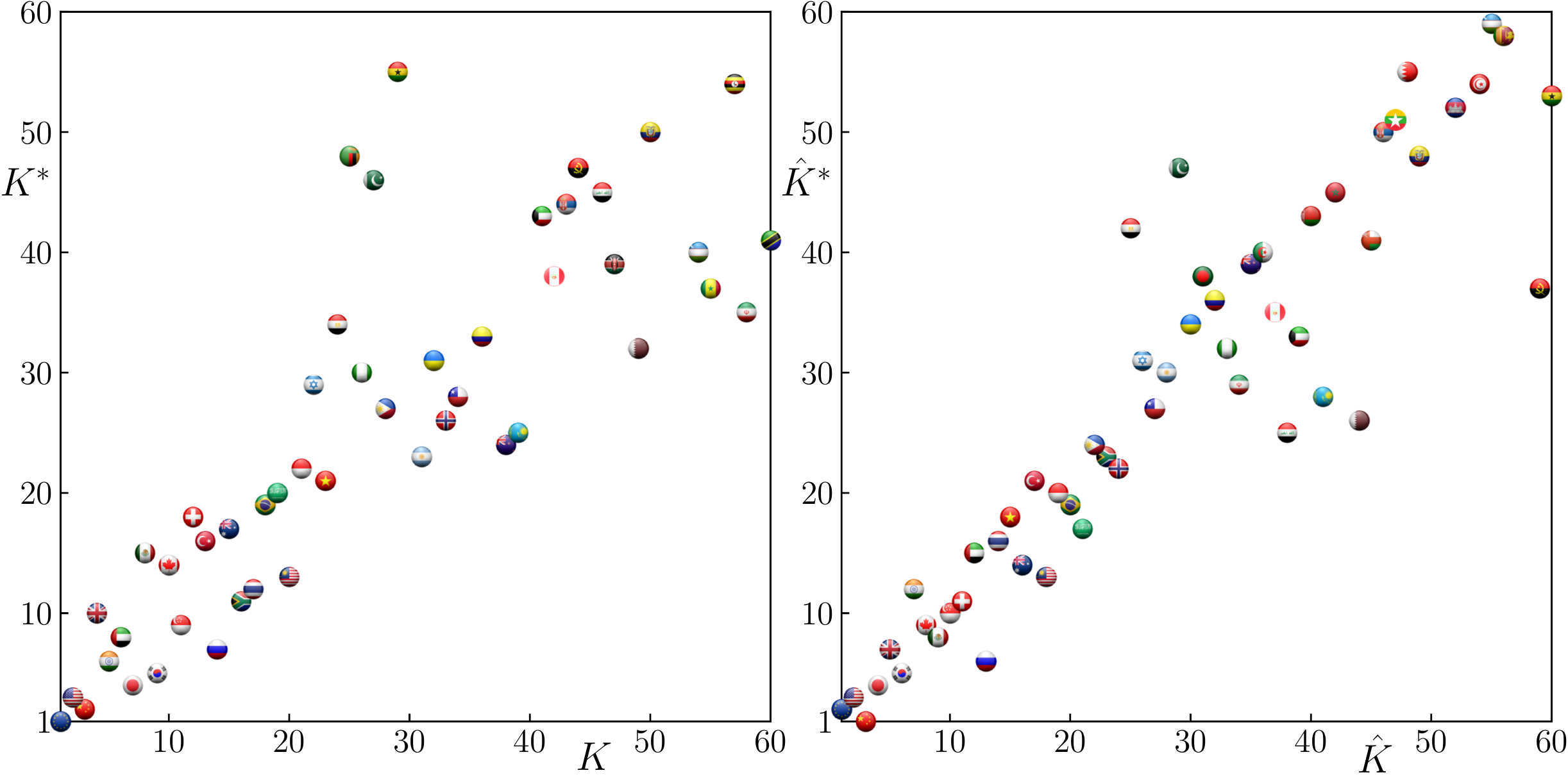}
	\caption{Circles with country flags show country positions on
  the plane of PageRank-CheiRank indexes $(K,K^*)$
  (summation is done over all products) (left panel)
  and on the plane of ImportRank-ExportRank  $\hat{K}$, $\hat{K}^*$ 
  from trade volume (right panel);
  data is shown only for index values less than $61$ in year 2018
  .} \label{fig1}
\end{figure}

As in \cite{wtn2,wtn3},
we define the trade balance of a given country
with PageRank and CheiRank probabilities given by
$B_c = (P^*_c - P_c)/(P^*_c + P_c)$.
Also we have from ImportRank and ExportRank probabilities
as $\hat{B}_c=  ({\hat{P}^*}_c - \hat{P}_c)/({\hat{P}^*}_c + \hat{P}_c)$.
The sensitivity of 
trade balance $B_c$ to the price of energy or machinery  can be obtained 
from the change of corresponding money volume
flow related to SITC Rev.1 code $p=3$ (mineral fuels) or $p=7$ (machinery)
by multiplying it by $(1+\delta)$, renormalizing column to unity and 
computing all rank probabilities and
the derivatives $dB_c/d\delta$. 

We also use the REGOMAX algorithm \cite{politwiki,wtn3}
to construct the reduced Google matrix $G_R$ 
for a selected subset of WTN nodes $N_r \ll N$.
This algorithm  takes into accounts all transitions of direct and
indirect pathways happening in the full Google matrix $G$ 
between $N_r$ nodes of interest. We use this $G_R$ matrix
to construct a reduced network of most strong transitions
(``network of friends'')
between a selection of nodes representing countries and products.

Even if Brexit enter into play in 2021, we use UN COMTRADE data of previous years
to make a projecting analysis of present and future power of EU composed of 27 countries.

Finally we note that GMA allows to obtain interesting results for 
various types of directed networks including Wikipedia  
\cite{geopolitics,celestin} and protein-protein interaction 
\cite{zinprotein,signor} networks.

\begin{table}[]
    \centering
    \caption{Top 20 countries of PageRank ($K$), CheiRank ($K^*$), 
    ImportRank and ExportRank in 2018.}
    \begin{tabular}{|c|l|l|l|l|}
    \hline
    Rank & PageRank                   & CheiRank                   & ImportRank           & ExportRank           \\ \hline
1    & EU                       & EU                       & EU                 & China                \\
2  & USA & China                                  & USA & EU                                  \\
3  & China                                  & USA & China                                  & USA \\
4    & United Kingdom             & Japan                      & Japan                & Japan                \\
5    & India                      & Repub Korea          & United Kingdom       & Repub Korea    \\
6    & U Arab Emirates       & India                      & Repub Korea    & Russia   \\
7    & Japan                      & Russia         & India                & United Kingdom       \\
8    & Mexico                     & U Arab Emirates       & Canada               & Mexico               \\
9    & Repub Korea          & Singapore                  & Mexico               & Canada               \\
10   & Canada                     & United Kingdom             & Singapore            & Singapore            \\
11 & Singapore                              & South Africa                           & Switzerland             & Switzerland             \\
12   & Switzerland & Thailand                   & U Arab Emirates & India                \\
13   & Turkey                     & Malaysia                   & Russia   & Malaysia             \\
14   & Russia         & Canada                     & Thailand             & Australia            \\
15   & Australia                  & Mexico                     & Viet Nam             & U Arab Emirates \\
16   & South Africa               & Turkey                     & Australia            & Thailand             \\
17   & Thailand                   & Australia                  & Turkey               & Saudi Arabia         \\
18   & Brazil                     & Switzerland & Malaysia             & Viet Nam             \\
19   & Saudi Arabia               & Brazil                     & Indonesia            & Brazil               \\
20   & Malaysia                   & Saudi Arabia               & Brazil               & Indonesia            \\ \hline
    \end{tabular}
\label{tab1}
\end{table}

\section{Results}\label{sec:results}
\subsection{CheiRank and PageRank of countries}

The positions of countries on the PageRank-CheiRank $(K,K^*)$ and
ImportRank-ExportRank $(\hat{K},\hat{K}^*)$ planes are shown in Fig.~\ref{fig1}
and in Table~\ref{tab1}.
These results show a significant difference between these two types of ranking.
Indeed, EU takes the top PageRank-CheiRank position $K=K^*=1$
while with Export-Import Ranking it has 
only $\hat{K}=1; \hat{K}^*=2$, with USA at $\hat{K}=2, \hat{K}^*=3$
and China at $\hat{K}=3, \hat{K}^*=1$. Thus EU takes the leading positions
in the GMA frame which takes into account  the muliplicity of trade transactions
and characterizes the robust features of EU trade relations.
Also GMA shows that UK position is significantly weakened compared to IEA description
(thus UK moves from $K^*=7$ in IEA to $K^*=10$ in GMA). From this data, we see also examples
of other countries that significantly improve there rank positions in GMA
frame compared to IEA: India ($K=5$, $K^*=6$, $\hat{K}=7$, $\hat{K}^*=12$), 
United Arab Emirates ($K=6$, $K^*=8$, $\hat{K}=12$, $\hat{K}^*=15$), 
South Africa ($K=16$, $K^*=11$, $\hat{K}=23$, $\hat{K}^*=23$). 
We attribute this to well developed, deep 
and broad trade network of these countries which are well captured by GMA in contrast
to IEA. Indeed, IEA only measures the volume of direct trade exchanges, while GMA characterises 
the multiplicity of trade chains in the world.

\begin{figure}[!h]
	\centering
	\includegraphics[width=\textwidth]{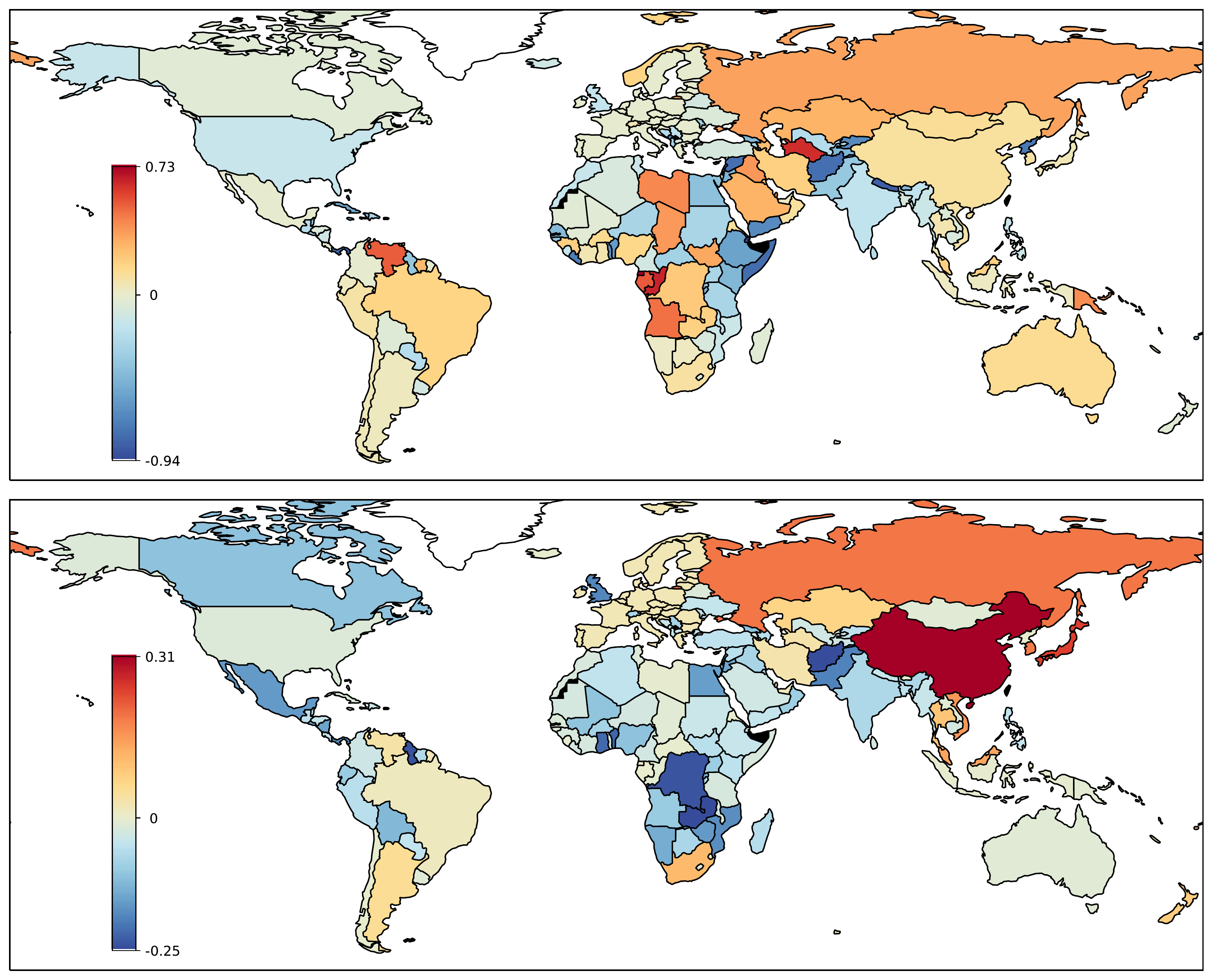}
	\caption{World map of trade balance of countries  
$B_c = ({P_c}^* - P_c)/({P_c}^* + P_c)$.
Top: trade balance values are computed from the trade volume of Export-Import;
bottom: trade balance values are computed from PageRank
and CheiRank vectors; $B_c$ values are marked by color with the corresponding 
color bar marked by $j$; 
countries absent in the UN COMTRADE report are marked by black color 
(here and in other Figs).}\label{fig2}
\end{figure}

\subsection{Trade balance and its sensitivity to product prices}

The trade balance of countries in IEA and GMA frames is shown in Fig.~\ref{fig2}.
The countries with 3 strongest positive balance are: 
Equatorial Guinea ($B_c=0.732$), Congo ($B_c=0.645$), Turkmenistan ($B_c=0.623$) in IEA and
China ($B_c=0.307$), Japan ($B_c= 0.244$), Russia ($B_c= 0.188$) in GMA.
We see that IEA marks top countries which have no significant world power
while GMA marks countries with real significant world influence.
For EU and UK we have  respectively $B_c=-0.015; 0.020$ (EU) and  
$B_c=-0.178; -0.187$ (UK) in IEA; GMA.
Thus the UK trade balance is significantly reduced in GMA
corresponding to a loss of network trade influence of UK
in agreement with data of Fig.~\ref{fig1} and Table~\ref{tab1}.
(We note that the balance variation bounds in GMA are smaller compared to IEA;
we attribute this to the fact of multiplicity of transactions 
in GMA that smooth various fluctuations which are  more typical for IEA).

\begin{figure}[t]
	\includegraphics[width=\textwidth]{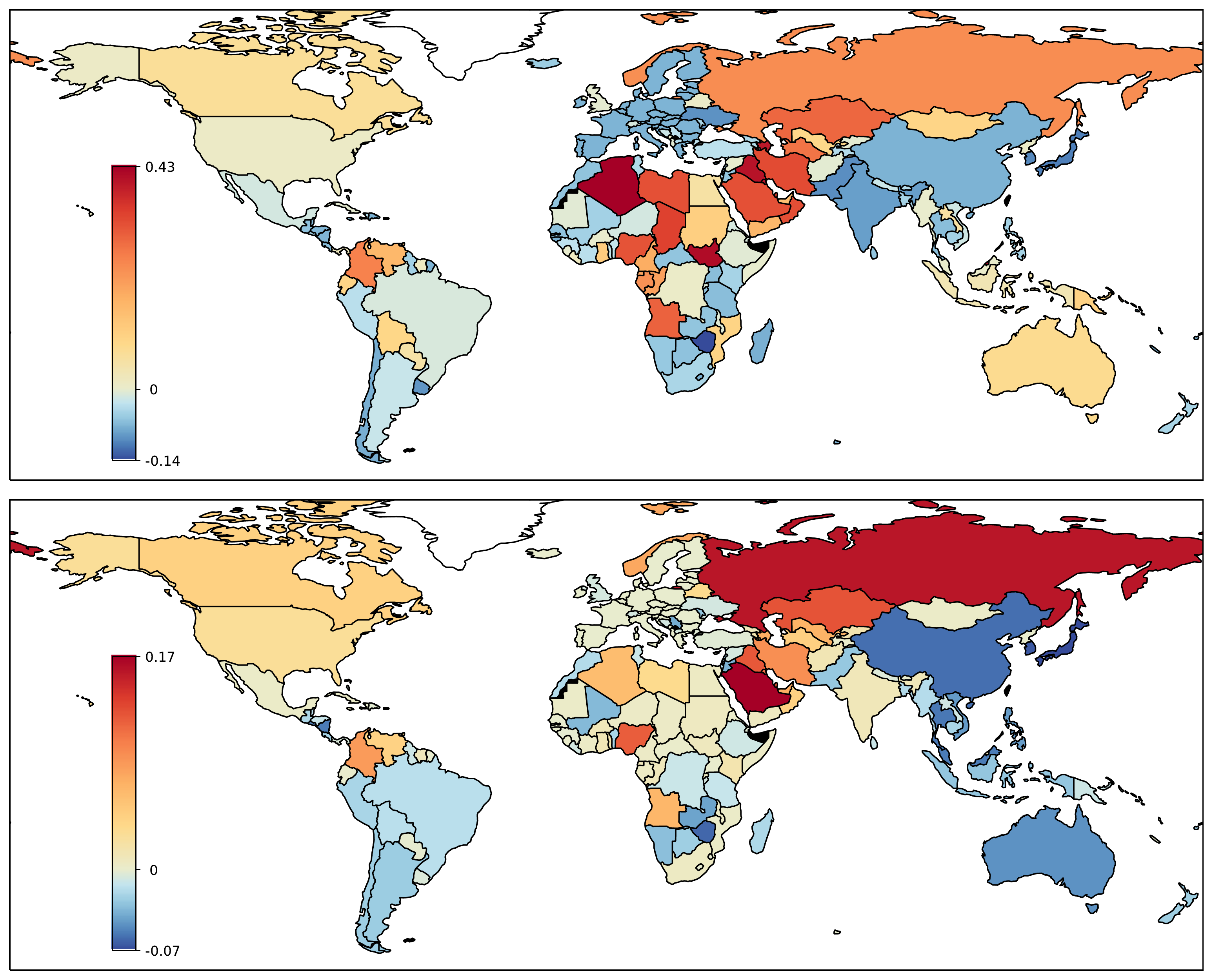}
	\caption{Sensitivity of country balance 
$dB_c/d\delta_s$ for product $s=3$ (mineral fuels).
Top: probabilities are from the trade volume of Export-Import;
bottom: probabilities are from PageRank
and CheiRank vectors. Color bar marked by $j$ 
gives sensitivity.} \label{fig3}
\end{figure}

The balance sensitivity $d B_c/d \delta_s$ to product $s=3$
(mineral fuels (with strong petroleum and gas contribution))
is shown in Fig.~\ref{fig3}. The top 3 strongest 
positive sensitivities $d B_c/d \delta_s$ 
are found for Algeria (0.431), Brunei (0.415), South Sudan (0.411) in IEA
and  Saudi Arabia (0.174), Russia (0.161), Kazakhstan (0.126)  in GMA.
The results of GMA are rather natural since Saudi Arabia, Russia and Kazakhstan are central petroleum producers. 
It is worth noting that GMA ranks Iraq at the 4th position. 
The 3 strongest negative sensitivities are Zimbabwe (-0.137), Nauru (-0.131), 
Japan (-0.106),  in IEA
and Japan (-0.066), Korea (-0.062), Zimbabwe (-0.058), in GMA.
For China, India we have $d B_c/d \delta_s$ values being respectively:
-0.073, -0.086 in IEA and -0.056, 0.010 in GMA. This shows that the trade network
of India is more stable to price variations of product $s=3$.
These results  demonstrate that GMA selects more globally influential countries.

\begin{figure}[t]
	\includegraphics[width=\textwidth]{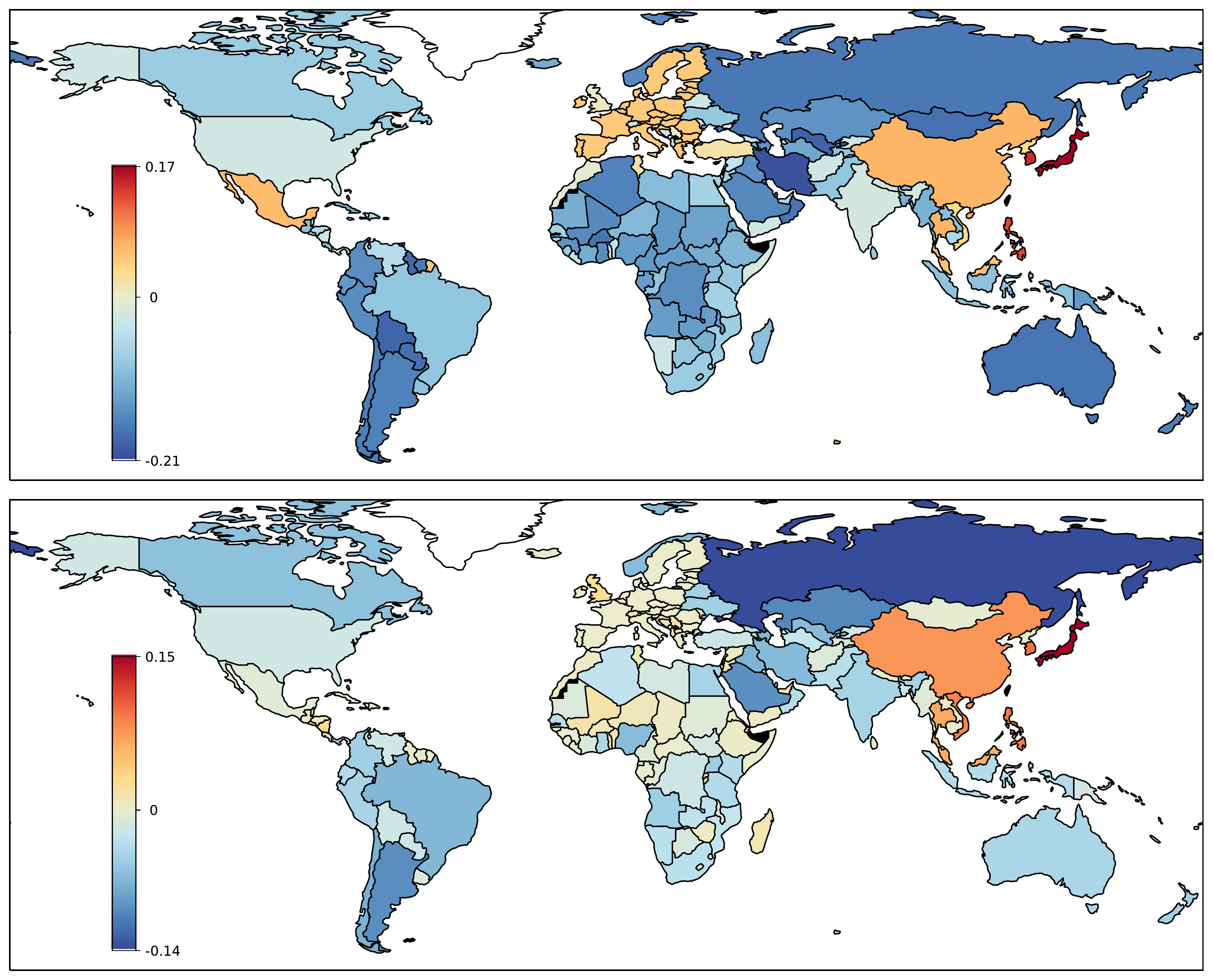}
	\caption{Same as in Fig.~\ref{fig3} but
for product $s=7$ (machinery).} \label{fig4}
\end{figure}

The balance sensitivity $d B_c/d \delta_s$ to product $s=7$
(machinery) is shown in Fig.~\ref{fig4}. 
Here the top 3 strongest 
positive sensitivities $d B_c/d \delta_s$ are found
in both  IEA and GMA for Japan (respectively 0.167, 0.151), 
Repub. Korea (0.143, 0.097), Philippines (0.130, 0.091).
The 3 strongest negative sensitivities are Brunei (-0.210), Iran (-0.202), 
Uzbekistan (-0.190)  in IEA
and Russia (-0.138), Kazakhstan (-0.102), Argentina (-0.097) in GMA.
Thus we again see that GMA selects more globally influential countries. 
The sensitivity $d B_c/d \delta_s$ values for EU, UK, China, Russia, USA are:
 EU (0.048), UK (0.006), China (0.065), Russia (-0.170), USA (-0.027) in IEA;
 EU (0.000), UK (0.024), China (0.077), Russia (-0.138), USA (-0.018) in GMA.
Latter GMA results show that even if machinery product ($s=7$) is very important for EU 
the network power of trade with this product becomes dominated by Asian countries
Japan, Repub. Korea, China, Philippines; 
in this aspect the position of UK is slightly better than EU. 

\begin{figure}[t]
	\includegraphics[width=\textwidth]{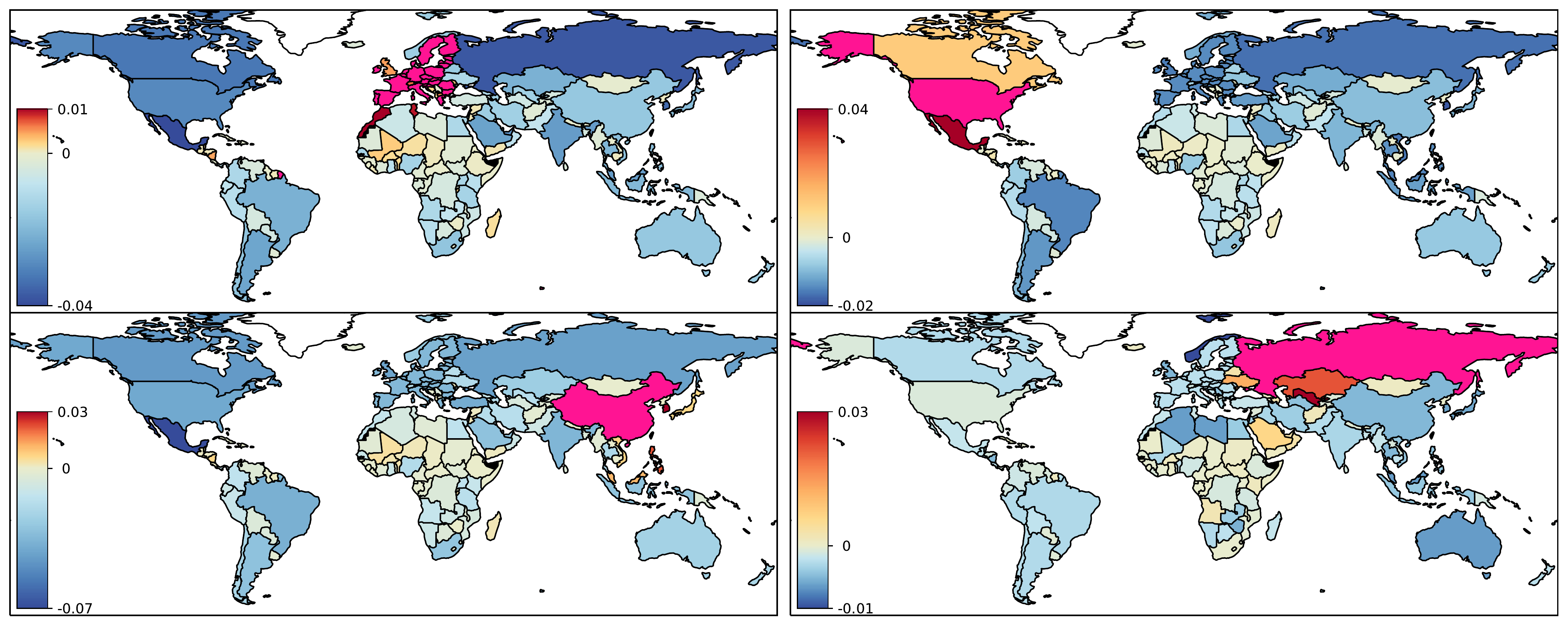}
	\caption{Sensitivity of country balance 
$dB_c/d\delta_{cs}$ for product price $s=7$ (machinery) 
of EU (top left), USA (top right), China (bottom left)
and $s=3$ (mineral fuel) of Russia (bottom right);
$B_c$ is computed from PageRank and CheiRank vectors;
sensitivity values are marked by color with the corresponding 
color bar marked by $j$. 
For EU, USA, China, Russia we have  $dB_c/d\delta_{cs} = 0.11, 0.11, 0.14, 0.12$
respectively, these values are 
marked by separate magenta color to highlight sensitivity 
of other countries in a better way.} \label{fig5}
\end{figure}

\begin{figure}[t]
	\includegraphics[width=\textwidth]{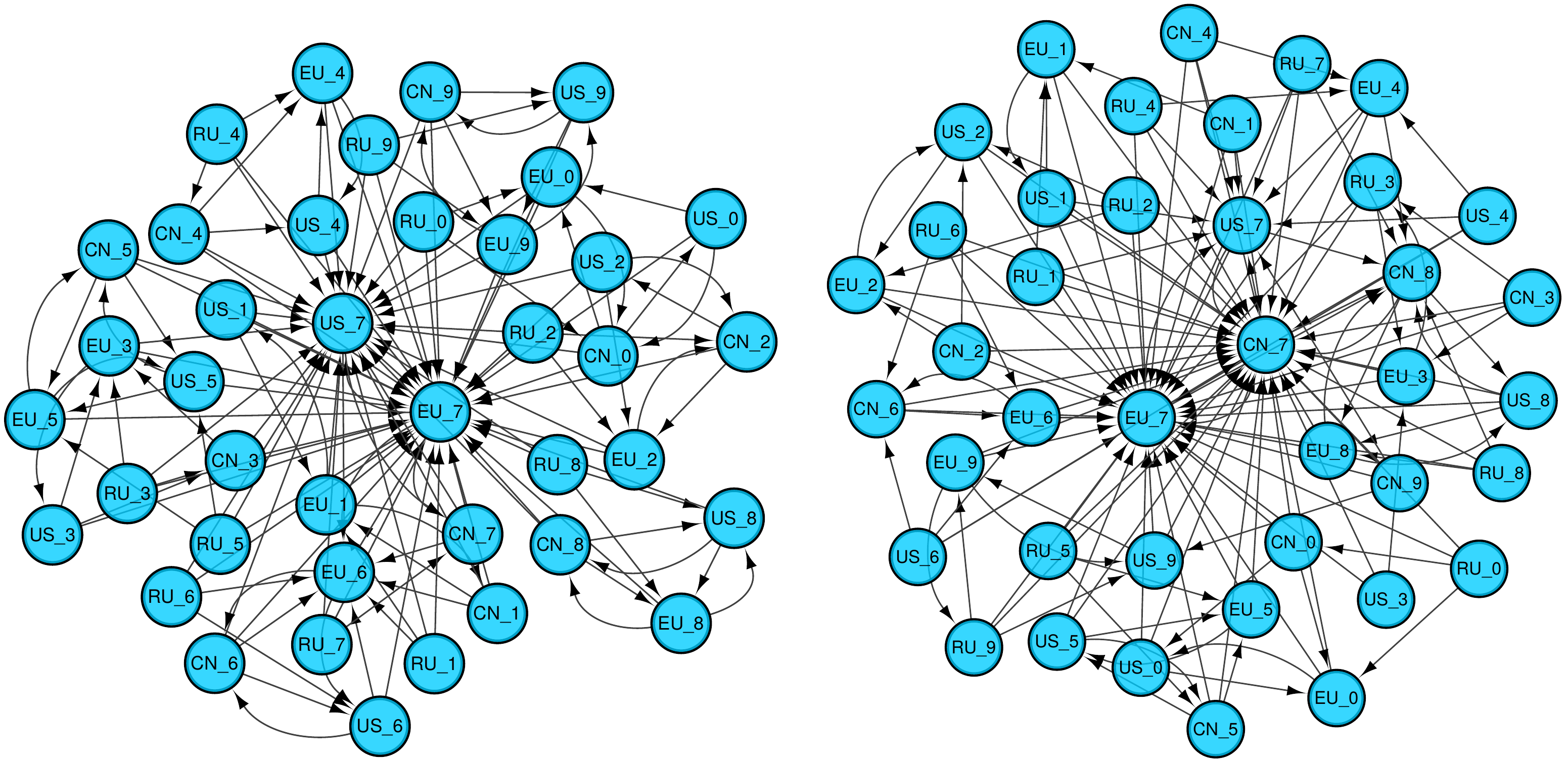}
	\caption{Network trade structure between 
EU, USA (US), China (CN), Russia (RU) 
with 10 products. 
Network is obtained from the reduced
Google matrix $G_R$ (left) and ${G^*}_R$ (right)
by tracing four strongest outgoing links
(similar to 4 ``best friends'').
Countries are shown by circles with two letters of country
and product index listed in Section 2. The arrow direction 
from node $A$ to node $B$ means that $B$ imports from $A$ (left)
and  $B$ exports to $A$ (right).
All $40$ nodes are shown.
} \label{fig6}
\end{figure}

In Figs.~\ref{fig3} and \ref{fig4}, we have considered the sensitivity of 
country balance to a global price of a specific product (mineral fuel $s=3$ or machinery $s=7$). 
In contrast, with GMA, we can also obtain the sensitivity of country balance to the price of
products originating from a specific country. Such results are shown in Fig.~\ref{fig5}.
They show that machinery ($s=7$) of EU gives a significant positive balance sensitivity 
for UK and negative for Russia. This indicates a strong dependence of Russia 
from EU machinery.   Machinery of USA gives strong positive effect for Mexico and Canada
with a negative effect for EU, Russia, Brazil, Argentina.
Machinery of China gives positive sensitivity for Asian countries
(Repub. Korea, Japan, Philippines) and significant negative effect for Mexico.
Mineral fuels ($s=3$) of Russia gives positive effect for Kazakhstan, Uzbekistan, Ukraine
(former USSR republics) and negative effect for competing petroleum and gas producers
Norway and Algeria.

\subsection{Network structure of trade from reduced Google matrix}

The network structure for 40 nodes of 10 products of EU, USA, China and Russia is 
shown in Fig.~\ref{fig6}. It is obtained from the reduced Google  matrix of $N_r=40$ nodes
of global WTN network with $N=1680$ nodes on the basis of REGOMAX algorithm
which takes into account all pathways between $N_r$ nodes via the global network of $N$ nodes.
The networks are shown for the direct  ($G$ matrix) and
inverted  ($G^*$ matrix) trade flows. For each node we show only 4 strongest outgoing 
links (trade matrix elements) that heuristically can be considered as the four ``best friends''. 
The resulting network structure clearly shows
the central dominant role of machinery product. For ingoing flows (import direction) of $G_R$ 
the central dominance of machinery for  USA and EU is directly visible while  
for outgoing flows (export direction), machinery of EU and China dominate exports.

It is interesting to note that the network influence of EU with 27 countries
is somewhat similar to the one constituted by a kernel of 9 dominant EU countries (KEU9)
(being Austria, Belgium, France, Germany, Italy,
Luxembourg, Netherlands, Portugal, Spain) discussed in \cite{keu9}.  This shows
the leading role played by these KEU9 countries in the world trade influence of EU.

Finally we note that additional data with figures and tables is available at \cite{ourwebpage}.

\section{Discussion}\label{sec:discussion}

We presented the Google matrix analysis of multiproduct WTN obtained from UN COMTRADE database
in recent years. In contrast to the legacy Import-Export characterization of trade,
this new approach captures multiplicity of trade transactions between world countries
and highlights in a better way the 
global significance and influence of trade relations between specific countries and products.
The Google matrix analysis clearly shows that the dominant position
in WTN is taken by the EU of 27 countries despite the leave of UK after Brexit.
This result demonstrates the robust structure of worldwide EU trade.
It is in contrast with the usual Import-Export analysis in which  USA and China
are considered as main players. We also see that machinery and mineral fuels
products play a dominant role in the international trade.
The Google matrix analysis stresses the growing dominance of 
machinery products of Asian countries (China, Japan, Republic of Korea).

We hope that the further development of Google matrix analysis 
of world trade will bring new insights in this complex system of world economy.

{\it Acknowledgments:}
We thank Leonardo Ermann for useful discussions.
This research has been partially supported through the grant
NANOX $N^\circ$ ANR-17-EURE-0009 (project MTDINA) in the frame 
of the {\it Programme des Investissements d'Avenir, France} and
in part by APR 2019 call of University of Toulouse and by 
Region Occitanie (project GoIA).
We thank UN COMTRADE for providing us a friendly access
to their detailed database.

%
%
%

\begin{thebibliography}{8}
%
%
%
%
\bibitem{eusite} {\it European Union},
        \url{https://europa.eu/european-union/about-eu/figures/economy_en#trade}
        (Accessed February (2021)).
\bibitem{wikibrexit} {\it Brexit},
        \url{https://en.wikipedia.org/wiki/Brexit}
        (Accessed February (2021)).
\bibitem{comtrade} {\it UN Comtrade database},
         \url{https://comtrade.un.org/}
        (Accessed February (2021)).
\bibitem{wtn1} Ermann L. and Shepelyansky D.L.:
        {\it Google matrix of the world trade network},
        Acta Physica Polonica A {\bf 120}, A158 (2011).
\bibitem{wtn2} Ermann L. and Shepelyansky D.L.: 
         {\it Google matrix analysis of the multiproduct world trade network},
         Eur. Phys. J. B {\bf 88}, 84 (2015).
\bibitem{wtn3} Coquide C., Ermann L., Lages J. and Shepelyansky D.L.:
          {\it Influence of petroleum and gas trade on EU economies from 
          the reduced Google matrix analysis of UN COMTRADE data},
          Eur. Phys. J. B {\bf 92}, 171 (2019).
\bibitem{keu9} Loye J., Ermann L. and Shepelyansky D.L.:
         {\it World impact of kernel European Union 9 countries
         from  Google matrix analysis of the world trade network},
          arXiv:2010.10962[cs.SI] (2020).
\bibitem{rmp2015} Ermann L., Frahm K.M. and Shepelyansky D.L.:
        {\it Google matrix analysis of directed networks},
          Rev. Mod. Phys. {\bf 87}, 1261 (2015).
\bibitem{politwiki} Frahm K.M., Jaffres-Runser K. and Shepelyansky D.L.:
         {\it Wikipedia mining of hidden links between political leaders},
          Eur. Phys. J. B {\bf 89}, 269 (2016).
\bibitem{brin} Brin S. and Page L.:
         {\it The anatomy of a large-scale hypertextual Web search engine},
         Computer Networks and ISDN Systems {\bf 30}, 107 (1998).
\bibitem{meyer} Langville A.M. and Meyer C.D.: 
        {\it  Google's PageRank and beyond: the science of search engine rankings}, 
        Princeton University Press, Princeton (2006).
\bibitem{ourwebpage} {\it Post-Brexit trade power of EU},
          \url{https://www.quantware.ups-tlse.fr/QWLIB/euwtn}
        (Accessed February (2021)).
\bibitem{geopolitics} Frahm K.M., El Zant S., Jaffres-Runser K. and Shepelyansky D.L.:
           {\it Multi-cultural Wikipedia mining of geopolitics interactions 
               leveraging reduced Google matrix analysis},
            Phys. Lett. A {\bf 381}, 2677 (2017).
\bibitem{celestin} Coquide C. and Lewoniewski W.:
         {\it Novel version of PageRank, CheiRank and 2DRank for Wikipedia in multilingual network 
          using social impact}, In: Abramowicz W., Klein G. (eds) 
         {\it Business Information Systems BIS}, Lecture Notes in 
         Business Information Processing {\bf 389}, 319 (2020).
\bibitem{zinprotein} Lages J., Shepelyansky D.L. and Zinovyev A.:
          {\it Inferring hidden causal relations between 
         pathway members using reduced Google matrix of directed biological networks},
          PLoS ONE {\bf 13(1)}, e0190812 (2018).
\bibitem{signor} Frahm K.M. and Shepelyansky D.L.:
            {\it Google matrix analysis of bi-functional 
         SIGNOR network of protein-protein interactions},
              Physica A {\bf 559}, 125019 (2020).

\end{thebibliography}
%

\end{document}


\title{Supplementary Material for: \\ Post-Brexit power of European Union \\ from the world trade network analysis}

\author{Justin Loye\inst{1,2} \and
Katia Jaffr\`es-Runser\inst{3}
\and
Dima L. Shepelyansky\inst{2}
}
\authorrunning{J.Loye et al.}
\titlerunning{Supplementary Material for: Post-Brexit power of European Union from the WTN}

\institute{Institut de Recherche en Informatique de Toulouse, \\
Universit\'e de Toulouse, UPS, 31062 Toulouse, France\\
\email{justin.loye@irit.fr}\\
\and
Laboratoire de Physique Th\'eorique, 
Universit\'e de Toulouse, CNRS, UPS, 31062 Toulouse, France\\
\email{dima@irsamc.ups-tlse.fr}\\
\and
Institut de Recherche en Informatique de Toulouse, \\
Universit\'e de Toulouse, Toulouse INP, 31071 Toulouse, France\\
\email{katia.jaffresrunser@irit.fr}\\
}
\maketitle  

\vspace{1.0cm}

\begin{figure}
    \centering
    \includegraphics[width=0.85\textwidth]{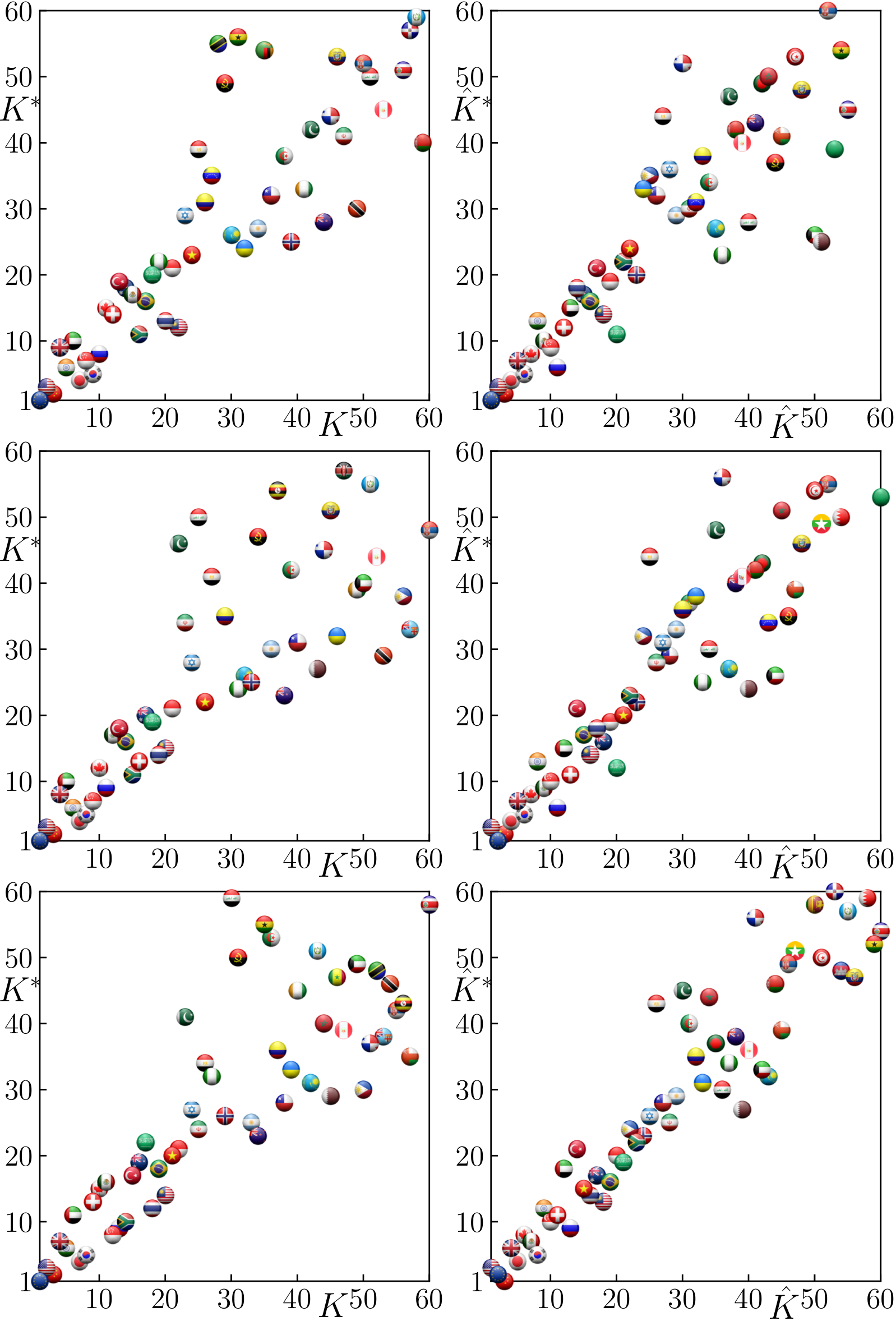}
    \caption{Circles with country flags show country positions on
    the plane of PageRank-CheiRank indexes $(K,K^*)$
    (summation is done over all products) (left panel)
    and on the plane of ImportRank-ExportRank  $\hat{K}$, $\hat{K}^*$ 
    from trade volume (right panel);
    data is shown only for index values less than $61$.
    First, second and third row correspond to years 2012, 2014, 2016.}
    \label{fig:KKstar_2012_2016}
\end{figure}

\begin{figure}
    \centering
    \includegraphics[width=0.96\textwidth]{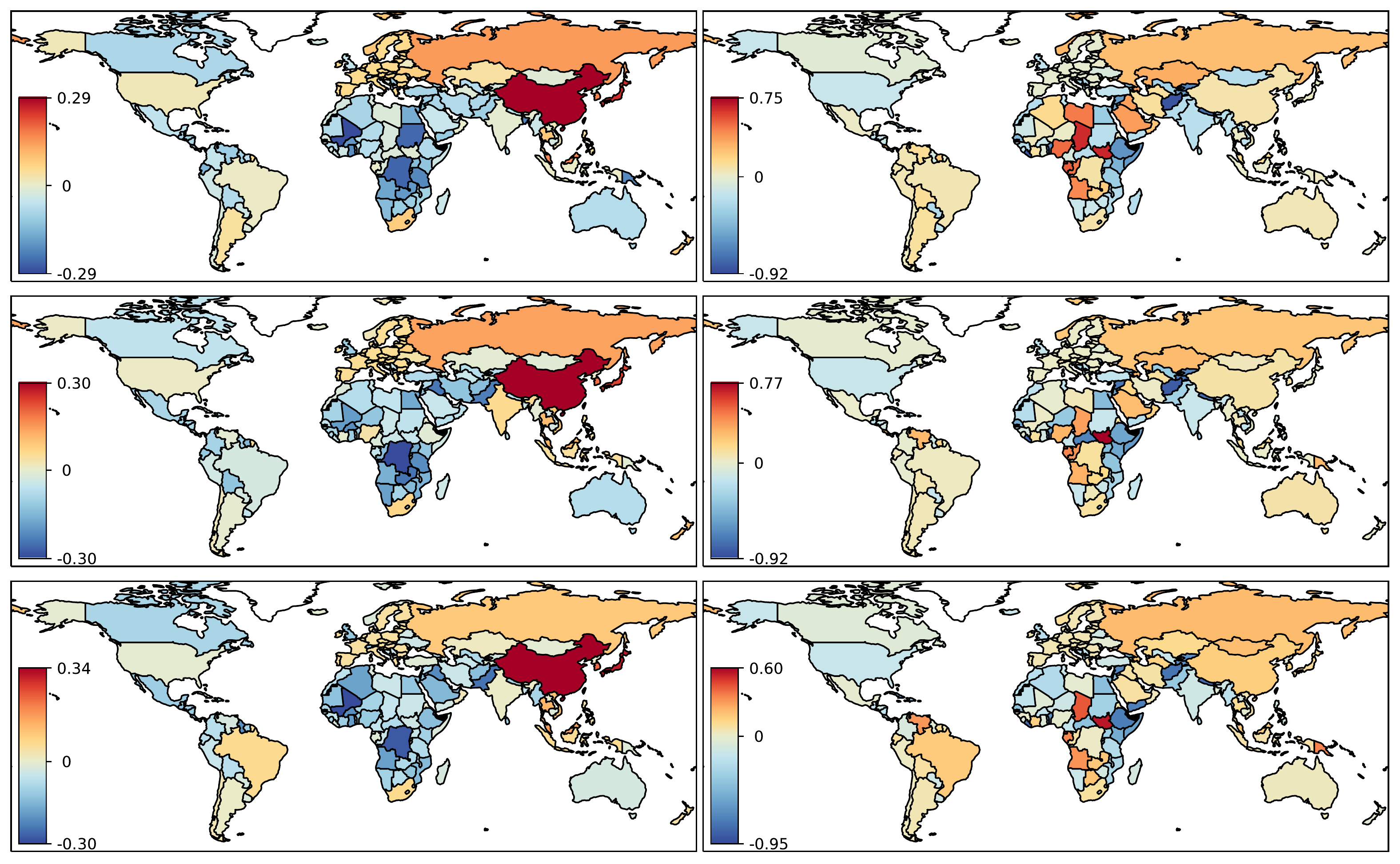}
    \caption{World map of trade balance of countries  
    $B_c = ({P_c}^* - P_c)/({P_c}^* + P_c)$.
    Left: probabilities are from PageRank and CheiRank vectors;
    Right: probabilities are from the trade volume of Export-Import;
    $B_c$ values are marked by color with the corresponding 
    color bar marked by $j$; 
    countries absent in the UN COMTRADE report are marked by black color 
    (here and in other SupFigs).
    First, second and third row correspond to year 2012, 2014, 2016.}
    \label{fig:balance2012_2016}
\end{figure}

\begin{figure}
    \centering
    \includegraphics[width=0.8\textwidth]{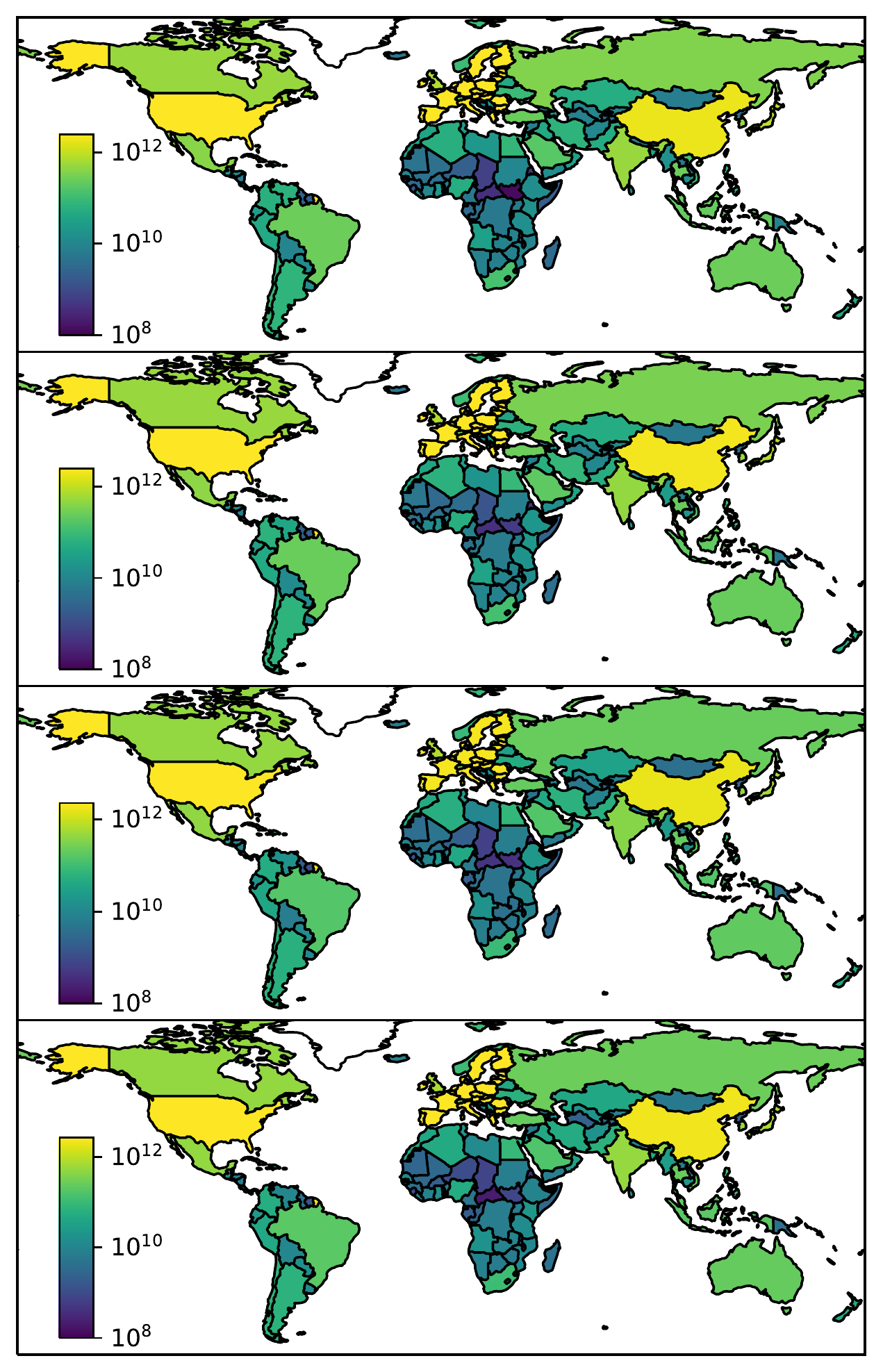}
    \caption{World map of countries importation volume in dollar.
    Values lower than $10^8$ are clipped for better visibility.
    First, second, third and fourth row correspond to years 2012, 2014, 2016, 2018.
    }
    \label{fig:importations}
\end{figure}

\begin{figure}
    \centering
    \includegraphics[width=0.8\textwidth]{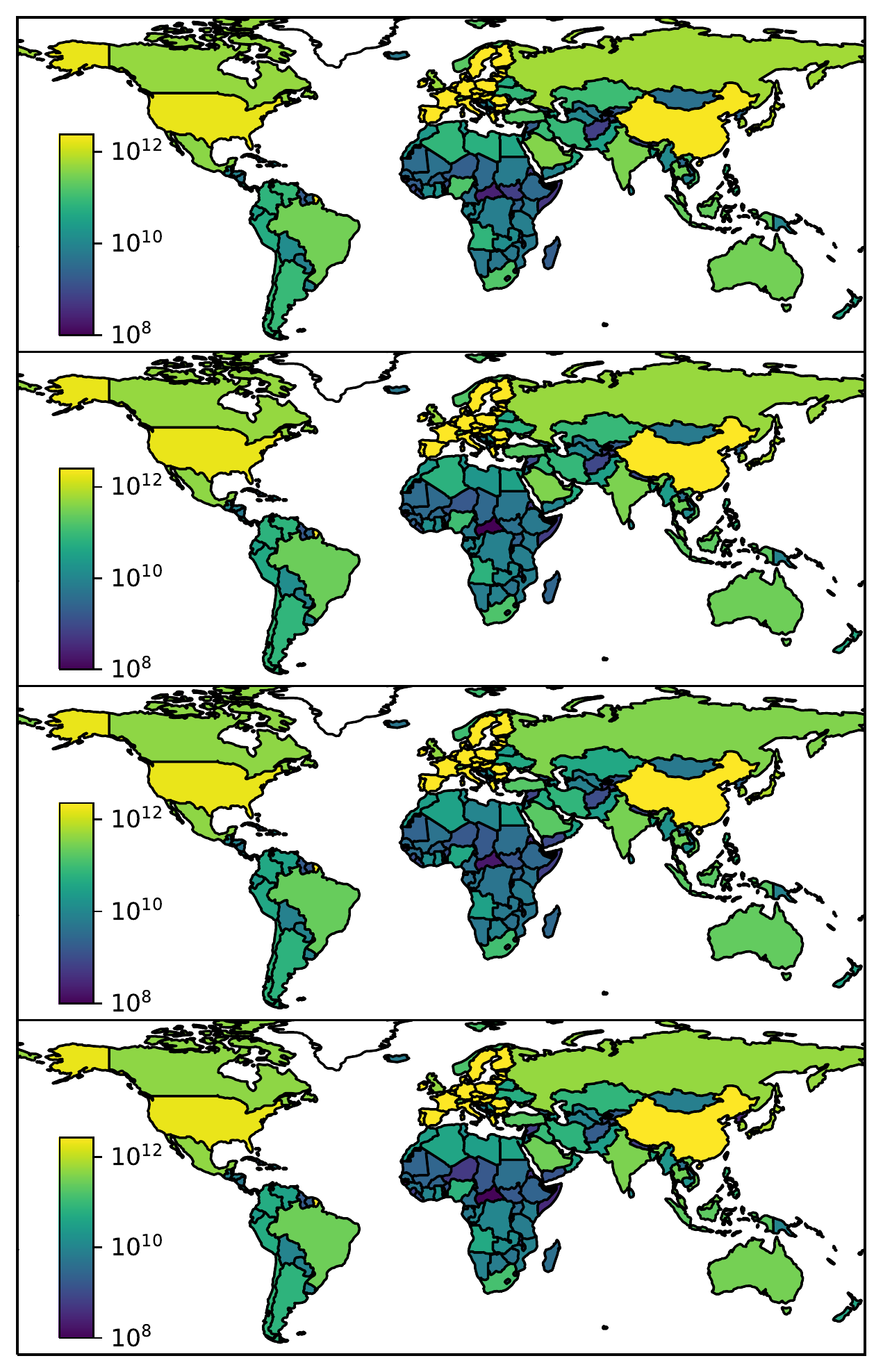}
    \caption{World map of countries exportation volume in dollar.
    Values lower than $10^8$ are clipped for better visibility.
    First, second, third and fourth row correspond to years 2012, 2014, 2016, 2018.
    }
    \label{fig:exportations}
\end{figure}

\begin{figure}
    \centering
    \includegraphics[width=0.96\textwidth]{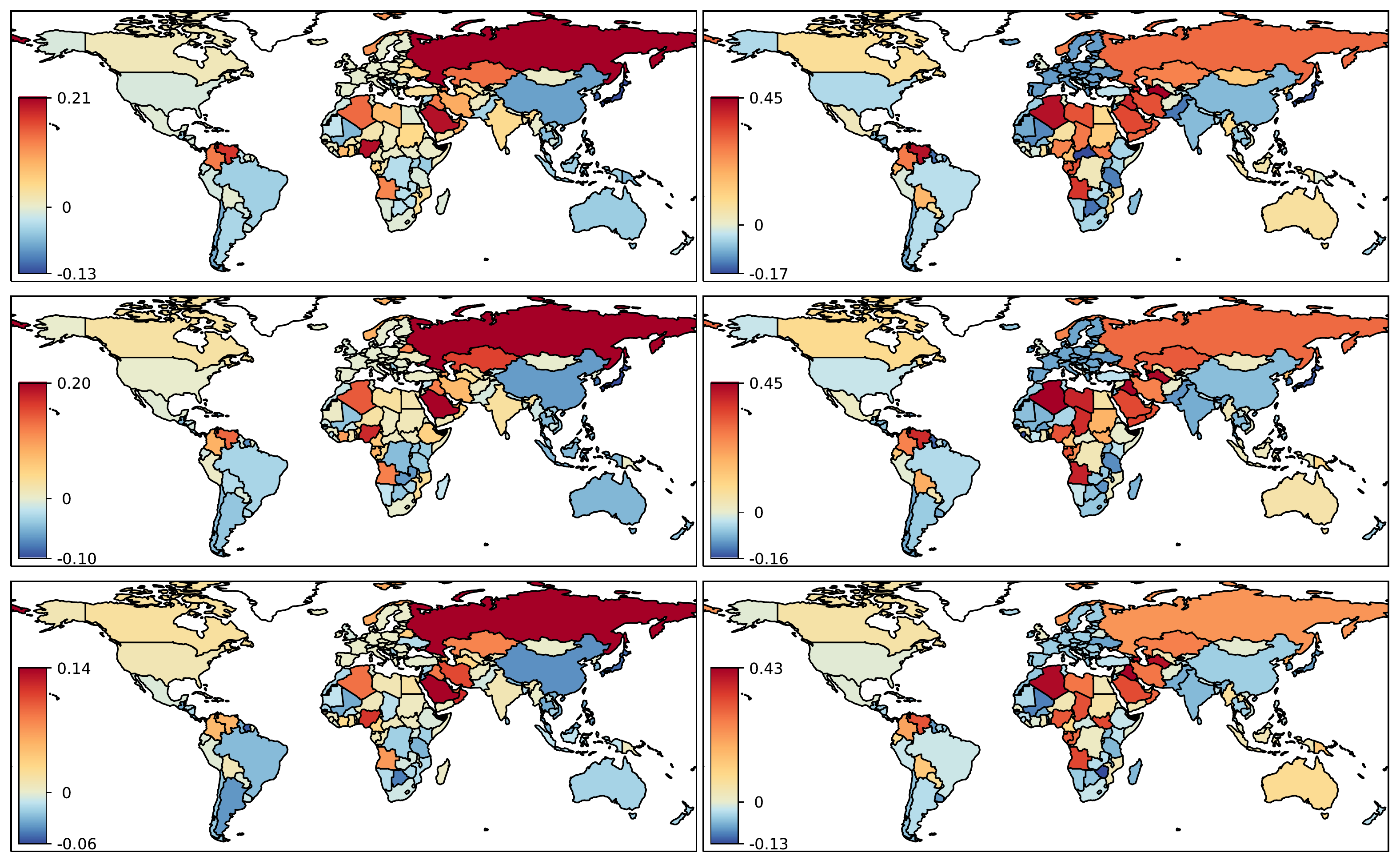}
    \caption{Sensitivity of country balance 
    $dB_c/d\delta_s$ for product $s=3$ (mineral fuels).
    Left: probabilities are from PageRank and CheiRank vectors;
    Right: probabilities are from the trade volume of Export-Import.
    Color bar marked by $j$ 
    gives sensitivity..
    First, second and third row correspond to year 2012, 2014, 2016.}
    \label{fig:dbalance_petroleum_2012_2016}
\end{figure}

\begin{figure}
    \centering
    \includegraphics[width=0.96\textwidth]{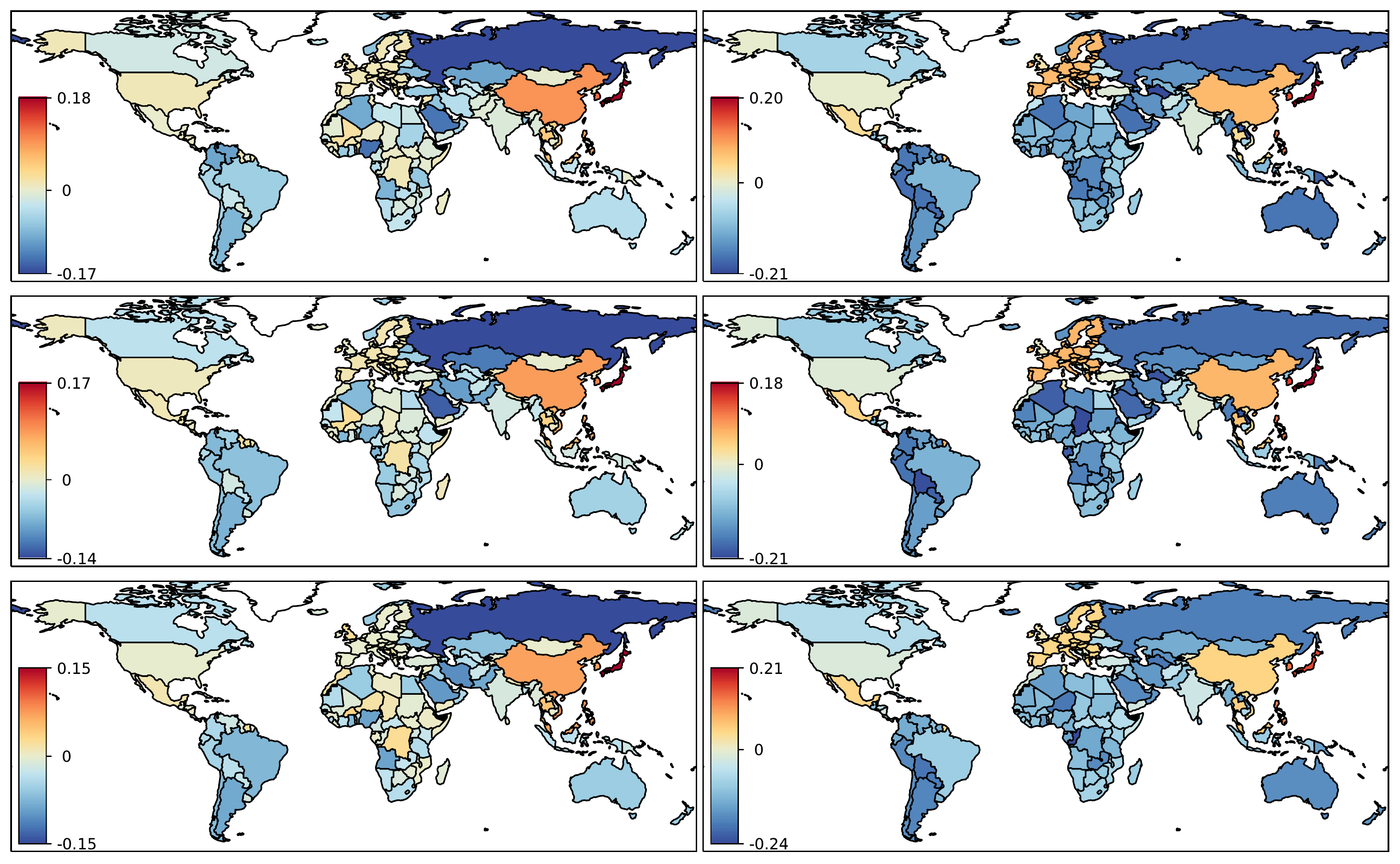}
    \caption{Same as in Fig.~\ref{fig:dbalance_petroleum_2012_2016} but
    for product $s=7$ (machinery).}
    \label{fig:dbalance_machinery_2012_2016}
\end{figure}

We present additional results SupFig.~\ref{fig:KKstar_2012_2016}, \ref{fig:balance2012_2016}, \ref{fig:importations},
\ref{fig:exportations}, \ref{fig:dbalance_petroleum_2012_2016} and \ref{fig:dbalance_machinery_2012_2016}.

SupFig.~\ref{fig:KKstar_2012_2016}, \ref{fig:balance2012_2016}, \ref{fig:dbalance_petroleum_2012_2016} and \ref{fig:dbalance_machinery_2012_2016}
reproduce for years 2012, 2014, 2016 respectively Fig.1, Fig.2, Fig.3 and Fig.4 of the main article. 
In SupFig.~\ref{fig:importations} and \ref{fig:exportations}, we respectively show the volume of importation and exportation of countries in the world trade network.



